# Large Language Models and Thematic Analysis: Human-AI Synergy in Researching Hate Speech on Social Media

**Petre Breazu** (University of Cambridge), **Miriam Schirmer** (Technical University of Munich), **Songbo Hu** (University of Cambridge), **Napoleon Katsos** (University of Cambridge)

**Abstract**

In the dynamic field of artificial intelligence (AI), the development and application of Large Language Models (LLMs) for text analysis are of significant academic interest. Despite the promising capabilities of various LLMs in conducting qualitative analysis, their use in the humanities and social sciences has not been thoroughly examined. This article contributes to the emerging literature on LLMs in qualitative analysis by documenting an experimental study involving GPT-4. The study focuses on performing thematic analysis (TA) using a YouTube dataset derived from an EU-funded project, which was previously analyzed by other researchers. This dataset is about the representation of Roma migrants in Sweden during 2016—a period marked by the aftermath of the 2015 refugee crisis and preceding the Swedish national elections in 2017. Our study seeks to understand the potential of combining human intelligence with AI's scalability and efficiency, examining the advantages and limitations of employing LLMs in qualitative research within the humanities and social sciences. Additionally, we discuss future directions for applying LLMs in these fields.

**Introduction**

In this article, we examine the capabilities of GPT-4 (OpenAI, 2024), the state-of-the-art Large Language Model (LLM) that powers ChatGPT, to perform a thematic analysis (TA) of YouTube comments related to the representation of Roma beggars in Sweden. The aim of this experiment is not to endorse LLMs for undertaking research tasks and making automatic decisions in relation to data analysis but rather to explore the advantages and limitations of a potential human-AI synergy to accelerate the analytical process.

TA is a well-established qualitative research method used across humanities and social sciences which is perfectly suited for innovative experiments with LLMs. Braun and Clarke (2006) outlined a clear methodology for TA which consists of six sequential steps: (i) familiarizing oneself with the data, (ii) generating initial codes, (iii) identifying themes, (iv) refining these themes, (v) defining and naming the themes, and (vi) preparing the final report. Traditionally, researchers have conducted both inductive and deductive TA. Inductive TA is a bottom-up, data-driven approach, according to which themes are derived directly from the data, without being influenced by the researcher's preconceptions or theoretical framework (Nowell, Norris et al. 2017). Conversely, deductive TA is a top-down, theory-driven approach, which starts with a predefined set of themes or theoretical framework that the researcher 'expects' to find in the data (Kennedy and Thornberg 2018). Both approaches are well-matched with the capabilities of LLMs. Inductive TA enables us to explore how LLMs independently identify themes directly from the dataset. This approach allows us to examine the models' ability to analyze data without predefined categories and input from researchers. Conversely, deductive TA offers an opportunity to observe how LLMs perform analysis within a structured framework in a more controlled setting. In this scenario, the researchers guide the model by introducing theoretical concepts and definitions of specific themes beforehand. This method facilitates a more directed analysis and shows how LLMs apply and adhere to predefined analytical criteria. In this study, we test the applicability of LLMs in supporting both inductive and deductive approaches to TA, with a view to explore the extent to which LLMs enhance the efficiency and comprehensiveness of qualitative research methodologies.

LLMs, as a category of Generative AI, are developed on an unprecedented scale in terms of model size (number of parameters), training data, and computational resources. For example, Meta AI's recent LLaMA-3 model is equipped with 70 billion parameters and was trained using 15 trillion tokens over 6.4 million Graphics Processing Unit (GPU) hours (Meta AI, 2024).

Such extensive training enables LLMs to efficiently perform a wide range of language-related tasks with zero-shot, one-shot, and few-shot learning[1], which require little or no task-specific data (Kaplan et al., 2020). The capability extends to both generative tasks such as text generation, translation, summarization, question answering, and dialog systems, as well as analytical tasks (often termed 'discriminative' in machine learning contexts) including sentiment analysis, named entity recognition, part of speech tagging, and text classification.

There is an emerging body of academic literature which shows how various LLMs have been used in qualitative research or display potential for performing some forms of TA (Dai, Xiong and Ku 2023, De Paoli 2023, Bano, Hoda et al. 2024). While LLMs have demonstrated remarkable proficiency in processing and understanding complex textual information (Bano, Zowghi and Whittle 2023, De Paoli 2023), their ability to effectively analyze and synthesize different types of data remains less explored. Previous studies on the use of LLMs in thematic analysis have employed a diverse array of data sources, including government reports (Khan et al., 2024), survey responses (Dai et al., 2024), semi-structured interviews (De Paoli, 2024), and legal documents, including criminal court opinions (Drápal et al., 2024). This variety of data types highlights the adaptability of LLMs across different fields and research contexts.

Our study contributes to the existing academic debates by focusing on a specific type of data: comments from social media platforms like YouTube, which are frequently laden with hate speech and inflammatory content (Breazu, 2023). This choice of dataset is significant for several reasons. Firstly, it represents a domain that has been less explored in existing research, particularly in the context of LLMs like GPT-4. Secondly, working with data that contains hate speech presents a unique challenge as it may not be processed by LLMs due to content policy restrictions. Such data, by its nature, often violates the guidelines set forth to ensure respectful and safe interactions within digital environments, and there is a risk that LLMs might not recognize the analysis of this data as a valid research task. This constraint is a critical area of concern in employing LLMs for research purposes. Our experimental study seeks to understand how to navigate these barriers responsibly. In this article, we examine GPT-4's capabilities to perform TA, identify potential contributions and limitations, and possibly provide a new perspective on future human-AI collaboration in academic research.

---

[1] Zero-shot learning refers to the ability of LLMs to complete tasks they have never been explicitly trained on, relying only on their pre-existing knowledge. One-shot learning refers to an LLM's ability to perform a task after receiving a single example or instruction, while few-shot learning adapts an LLM to a new task by being exposed to only a minimal number of specific examples. These methods, in contrast to fine-tuning, highlight the versatility and generalisation abilities of LLMs. They allow LLMs to complete a wide range of tasks without extensive task-specific training datasets, which typically require thousands of examples, to reach optimal performance.

**LLMs and Qualitative Research**

The integration of LLMs in qualitative research presents both promising opportunities and challenges (Dai, et al., 2023; De Paoli, 2023). In what follows we highlight recent academic debates which addressed the potential role of LLMs in enhancing qualitative research methodologies. Most academic literature located so far, particularly focused on TA and the complex interplay between human researchers and AI technologies.

Although the deployment of LLMs in qualitative research is in 'its status nascendi' [in the state of being born] (De Paoli, 2023:3), there are ongoing debates about their role and effectiveness in processing and analyzing data. Some researchers (Byun et al 2023; Rietz and Maedche, 2021) suggest that AI can match human capabilities in processing qualitative data and highlight the potential for LLMs to learn human coding practices which suggest that these models can adapt to the subjective nature of the qualitative analysis. Byun et al. (2023) argue that LLMs could rival human capabilities in generating and analyzing qualitative content, especially because of their ability to process large amounts of data and expedite the analysis. Researchers also highlight the potential of LLMs to overcome typical limitations of qualitative research performed by human researchers, especially in relation to processing large datasets, generalizing results to larger contexts, and avoiding subjectivity.

Other studies (Bano et. al 2024; Rudolph et al. 2023) caution against over-reliance on LLMs, pointing out discrepancies between AI and human reasoning that could affect the interpretation of qualitative data. These authors also point to the limitations of LLMs, especially in relation to fully understanding the context of research or the complex nature of human communication. It remains unclear how LLMs compare to human intelligence when performing various qualitative analytical tasks. Bano et al. (2023) and Rudolph et al. (2023) also draw attention to the risks of 'hallucinations' — instances where LLMs generate inaccurate or fabricated information. These inaccuracies, alongside the issue of 'model drift'[2] and the limitations imposed by LLMs' inability to access or interpret the full breadth of relevant literature, present significant difficulties to the validity and replicability of research findings. Furthermore, the evolving nature of copyright and intellectual property concerns (Balel, 2023; Polonsky and Rotman, 2023) needs a thoughtful approach to the integration of LLMs in academic work [here we refer to LLM as a co-author].

---

[2] Model drift refers to the degradation of model performance over time due to changes in the underlying data distribution or the relationship between input features and the target variable, which can result in reduced accuracy and reliability of predictions.

It is undeniable that LLMs offer unparalleled advantages in processing large datasets and significantly streamline the analysis process. The ongoing debates surrounding their application in qualitative research point to a delicate balance between embracing technological advances and exercising prudence. While acknowledging that LLMs augment our research capabilities with their speed and scale, AI should complement rather than substitute the critical insight that only human expertise can provide (De Paoli, 2023; Gao et al. 2023).

**Data and Context of Research**

This article uses data from an EU-funded research project that explores the representation of Roma in Swedish media and political discourse, focusing specifically on Romaphobia as evidenced in comments on YouTube videos about Roma beggars in Sweden. Following an analysis of Roma beggars' portrayal in four leading Swedish newspapers (Breazu, 2024; Breazu and Machin 2024), this data set aims to understand how such discourses resonate or not with the general public on social media. For this experiment, we selected a set of 474 YouTube comments which were thematically categorized by an early career researcher, using NVivo[3].

The socio-political backdrop of this research is crucial for contextual understanding. In Sweden, begging is legally considered a form of free expression and is protected by the Constitution. The 2007 EU enlargement, which saw Romania and Bulgaria's accession, led to many migrants, including ethnic Roma, into Sweden, drawn by the promise of better economic prospects (Breazu, 2024). However, challenges such as limited education and language barriers left some Roma migrants unable to find work or housing, pushing them towards begging or busking. The consequent visibility of Roma begging in public areas sparked debates on public order, safety, and well-being which led to an increase in anti-Roma sentiments (Hansson, 2023; Wigerfelt and Wigerfelt, 2015).

Throughout the years, discussions on potentially banning begging have surfaced repeatedly. The refugee crisis in 2015 notably shifted public and political discourse, intensifying debates around begging bans. By 2016, amidst rising political attention, especially before the 2017 elections, the dilemma of Roma begging and the prospect of instituting localized bans emerged as significant issues in Swedish politics. The analysis of these data sets seeks to offer insights into how these debates are taken up by social media users in their online engagement.

---

[3] NVivo is a qualitative data analysis (QDA) software that helps researchers organize, analyze, and find insights in unstructured or qualitative data such as interviews, open-ended survey responses, articles, social media, and web content.

**Experiment Design**

For our experimental design, we employed a two-fold approach using OpenAI's GPT-4 architecture via the OpenAI API. First, GPT-4 was given various segments of the dataset containing YouTube comments about Roma migrants in Sweden and was tasked to inductively categorize these comments. We assigned ChatGPT-4 the role of a researcher and tasked it to follow Braun and Clarke's (2006) six steps of conducting a thematic analysis on our YouTube dataset about the representation of Eastern European Roma beggars in Sweden. The steps included initial reading of data, coding the data by highlighting key phrases or sentiments, identifying overarching themes based on the coded data, and providing brief descriptions for each identified theme. We fed the dataset in seven separate batches, and we allowed the model to independently analyze the comments without providing pre-defined categories or theoretical framework to ensure an organic emergence of themes based solely on ChatGPT-s's reading of the comments. This thorough approach allowed us to assess the thematic classification capabilities of GPT-4 and gain insights into its alignment with human evaluators and its overall efficacy in qualitative research tasks. These categories identified by GPT-4 were then compared with those found by a human qualitative researcher. Additionally, four more experts in qualitative thematic analysis who were familiar with the dataset assessed the quality of the categories.

Second, we used the identified categories to instruct GPT-4 to deductively assign each comment to one of the previously established categories. We used the OpenAI GPT-4 API with a temperature setting of 0.1 for all API analyses. Throughout this process, we experimented with multiple variations of prompts to optimize our results. We started with a basic prompt employing role-prompting, a fundamental technique in prompt engineering (Chen, Zhang, Langrené, & Zhu, 2023). Assigning the model a specific role, such as an expert, has been proven to be more effective in guiding the model's responses. In our prompts, we assigned the model the role of a 'qualitative researcher investigating the representation of Roma in YouTube comments.' The initial prompt included a basic task description and a short description of the categories. Following best practices in prompt engineering (Chen et al., 2023; Hu et al., 2024;

Liu et al., 2023), we progressively enriched the prompt with additional information and instructions. Mu et al. (2023) demonstrated that augmenting GPT prompts with detailed task and label descriptions significantly boosts its performance. To further refine the prompt, we manually reviewed randomly selected comments and their assigned labels. During this process, experts agreed that some of the comments, such as the one discussed above, did not fit any of the categories found by GPT-4. For example, the comment *"'facts'" that everyone can see are usually the wrong "facts". Facts are and can be verified, not just something one person makes up as he goes."* This was labeled as "Ethnic Misunderstanding" although there was no clear hint on who the comment was targeted at or what it was about in detail. Recognizing this limitation, we added a "None" option to the original version of the prompt, allowing for responses that didn't neatly fit the existing categories. This step was important for identifying discrepancies between the model's classification and expert evaluations, as well as for detecting patterns in any misclassifications.

To obtain a more generalizable comparison of GPT's categorization quality, we conducted the deductive analysis twice using different sets of categories. In the first instance, as has been described in this section, the categories were created through the ChatGPT interface. In the second instance, they were assembled by a qualitative researcher familiar with the field.

**Findings and Discussions**

The initial observation is that, beyond its high level of efficiency and ability to process large data sets in seconds, GPT-4 follows steps similar to those of human researchers.

*Initial ChatGPT Categorization Scheme*

Leveraging its neural architecture and attention mechanism, GPT-4 can attend to every word in the comments, and the initial analysis resulted in the identification of 152 themes. This mirrors the experience of human researchers, where an initial analysis often produces a multitude of themes. Subsequent steps involve instructing GPT-4 to eliminate redundancy, statistically insignificant themes, and overlapping repetitions and generate a refined set of themes. By applying the same process, GPT-4 ultimately distilled the data into five main categories, as illustrated in Table 1.

In our initial round of categorizing YouTube comments, GPT-4 identified the following categories that best describe the dataset after analyzing 474 posts in 7 batches: Ethnic Misunderstanding and Identity Confusion, Stereotyping and Social Prejudice, Economic

Concerns and Welfare Debates, Cultural Clash and Integration Challenges, and Polarization of Public Opinion (Table 1).

**Table 1: ChatGPT categories to describe the dataset.**

| Category | Description |
|---|---|
| Ethnic Misunderstandings and Identity Confusion: | Emphasizes the distinction between Roma individuals and the ethnic majorities of Romania and Bulgaria, urging accurate ethnic identification and addressing misconceptions about nationality versus ethnicity. Discusses the Roma's historical migration from India, emphasizing their unique cultural evolution in Europe, the complexity of their identity, and the challenges in categorizing Roma strictly based on their ancient origins. |
| Stereotyping and Social Prejudice: | Stereotypes and prejudices evident in public discourse about begging, crime, theft; and the impact of stereotypes on national reputations. Concerns about crime, exploitation, and the association of Roma with organized criminal activities, including aggressive begging and scams. |
| Economic Concerns and Welfare Debates: | Explores the economic implications of migration, public perceptions of immigrants and minorities, and critiques of current policies affecting societal integration, welfare systems, and public services. |
| Cultural Clash and Integration Challenges: | Highlights the differences in cultural norms and legal adherence between Roma communities and the broader populations, touching on the perception of separate legal systems or tribal laws within Roma communities. |
| Polarization of Public Opinion: | Engages in a broader debate on what constitutes national identity and citizenship in the context of global migration, including discussions on multiculturalism, societal change, and the preservation of cultural identity amidst demographic shifts. |

The initial classification performed by GPT-4 on a dataset of YouTube comments about Roma in Sweden yielded categories that aligned well with those identified in an earlier thematic analysis by a qualitative researcher. These categories were reviewed to ensure they adequately captured the scope and addressed every aspect of the comments. To validate the accuracy and relevance of the categories produced by GPT-4, four domain experts independently compared the results. Upon review, the experts agreed that the categories made sense and were comparable to the established thematic framework and that all relevant aspects of the comments were appropriately addressed.

We will now examine how GPT-4's thematic insights into YouTube comments about Eastern European Roma beggars in Sweden compare to the ones by the human researcher.

Both analyses identified the theme of misconceptions surrounding the ethnic identity of the Roma people. GPT-4 categorizes this under 'Ethnic Misunderstandings and Identity Confusion,' and frames it around the historical migration of Roma from India and the complexity of Roma identity. In contrast, the human researcher's category, '(Non)Belonging,' emphasizes a common discriminatory public discourse about Roma as the 'other' European, which although live in Europe are not part of the nation (Marin Thornton, 2014; McGarry, 2014) This complex view reflects the human researcher's reliance on academic literature on the discursive representations of Roma and racism to capture the layers of identity politics.

The prevalence of negative stereotypes and prejudices against Roma is another shared theme. GPT-4's category, 'Stereotyping and Social Prejudice, discusses the impact of these stereotypes on specific national reputations and the association with criminal activities. Meanwhile, the human researcher identifies 'Perceptions and Stereotypes,' which focuses specifically on the depiction of Roma as beggars, thieves, and unproductive citizens. The human researcher's detailed focus underscores the informed understanding of stereotypes' roots and impacts, influenced by scholarly insights (Rosenhaft and Sierra, 2000, Tremlett, 2022, van Dijk, 2000)

Economic implications and the strain on welfare systems feature prominently in both analyses. GPT-4's 'Economic Concerns and Welfare Debates' captures public perceptions of Roma as economic migrants exploiting welfare policies. In contrast, the human researcher categorizes this discourse under 'Populism' and 'Nativism.' These terms reflect a more precise academic framing of how economic concerns intersect with political narratives about immigration, national identity, and cultural threats, drawing on established theories in political science and sociology (Betz, 2019, Krzyżanowski et. al, 2021; Newth, 2023).

Both analyses highlight the cultural differences and integration challenges faced by Roma communities. GPT-4 uses the category 'Cultural Clash and Integration Challenges' to discuss perceived legal and cultural separations. The human researcher's 'Cultural Racism' points to racism expressed through cultural markers of otherness, rather than biological categories. This distinction is informed by academic discussions on modern forms of racism that focus on cultural incompatibility instead of overt biological racism (Bonilla-Silva, 2013; Breazu and Machin, 2024).

Finally, both analyses address polarized public opinions. GPT-4's 'Polarization of Public Opinion' discusses broader debates on national identity and multiculturalism, capturing the societal divide. The human researcher, however, identifies 'Extreme Hate Speech,' clearly identifying various forms of abusive and dehumanizing language that endorses violence. This specificity is informed by existing academic research on hate speech (Guiora and Park, 2017; Matamoros-Fernández and Farkas, 2021). Additionally, GPT-4's focus on neutral language could also contribute to its broader categorization, as the model avoids to explicitly label the comments as hate speech.

*Potential Causes for Differences*

The differences between GPT-4 and human researcher analyses can be attributed to several factors. GPT-4's approach tends to categorize themes in a broad, generalized manner, focusing on overarching social and cultural issues. This is likely due to its design as an AI model trained to process and summarize vast amounts of text without the depth of specialized academic training. Consequently, its analysis offers a wide lens on the topics, suitable for capturing a broad spectrum of public discourse.

On the other hand, the human researcher's analysis is deeply informed by existing academic literature, which provides a more detailed understanding of the issues. The use of specific terms like 'Populism,' 'Nativism,' and 'Extreme Hate Speech' reflects a thorough grounding in scholarly work on the representation of Roma migrants. This specificity is essential in identifying the subtle manifestations of racism and discrimination, which may not be as readily apparent in a more generalized analysis.

Moreover, the human researcher's focus on socio-political narratives and the role of media and political elites demonstrates an understanding of the broader context in which these public opinions are formed. This perspective is crucial for comprehending how public discourse is shaped and the implications it has for societal attitudes and policies.

When evaluating the classification, experts agreed that some of the comments did not fit neatly into the categories identified by GPT-4. For example, comments specifically targeting Roma in a derogatory manner were usually labeled as 'Ethnic Misunderstandings and Identity Confusion' or 'Stereotyping and Social Prejudice.' This is one example:

*with the lack of Gypsies in Bulgaria, property prices have been on the rise lately. Keep the Gypsies ... send down the sexy Swedish bikini team to Sunny Beach Bulgaria, where blonds are welcome and Gypsies are not"*

The human researcher labeled this comment as hate speech due to its dehumanizing language which reinforces harmful stereotypes (e.g. the use of the pejorative *Gypsies*[4]), and explicitly endorses the exclusion of Roma individuals based on racial prejudice (*where blonds are welcome and Gypsies are no*t).

Recognizing this limitation, a 'None' option was added to the categorization process to allow for responses that didn't neatly fit the existing categories. The addition of the option not to assign a label to a comment proved to serve multiple purposes: it enhanced the accuracy of categorization by preventing misclassification, revealed potential gaps in the thematic framework by highlighting the proportion of responses that fall outside predefined categories, reduced bias by avoiding forced categorization, and acknowledged the complex nature of immigration discussions that may not be easily captured by broad themes.

Figure 1 illustrates the distribution of themes related to immigration discussions as categorized by GPT-4, comparing scenarios with and without a 'None' option. The data reveals that 'Stereotyping and Social Prejudice' is the most prevalent theme when specific categories are required, followed by 'Polarization and Public Opinion' and 'Ethnic Misunderstandings and Identity Confusion' However, when a 'None' option is introduced, it becomes the most frequently selected choice (193 samples; 40.72% overall), surpassing all other categories. This shift suggests that many responses do not neatly fit into the predefined themes, highlighting the complexity of immigration discourse. The introduction of the 'None' option also leads to a decrease in the frequency of all other themes, indicating that forced categorization may overestimate the prevalence of certain topics.

Notably, 'Cultural Clash and Integration Challenges' remains the least common theme in both scenarios. Overall, this visualization illustrates the complex nature of immigration discourse and the importance of flexible categorization in capturing the full range of perspectives.

---

[4] The term 'Gypsy' in the Eastern European context is considered derogatory. It does not have the semantic value to accurately reflect the ethnic identity of the Roma people but rather carries negative connotations such as being unreliable, lazy, dirty, quarrelsome, or deviant. It is recommended to use 'Roma' or 'Romani' to refer to this ethnic minority respectfully.

**Figure 1: GPT-4 Theme Distribution (GPT-Themes)**

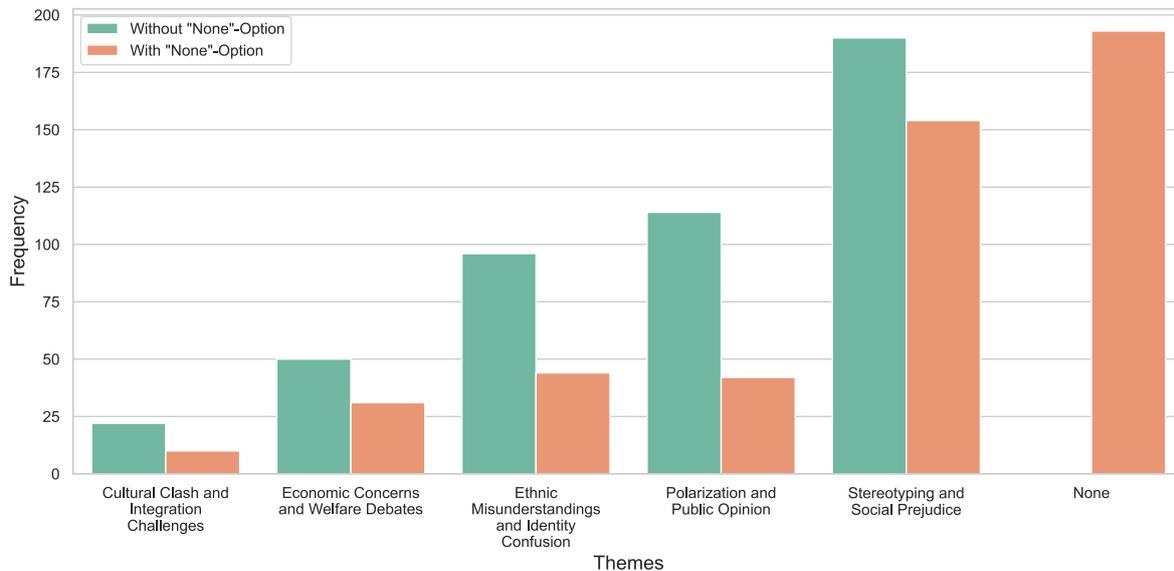

**Human-Supported Categorization**

Unlike GPT-4, the human researchers identified more specialized labels, such as belonging, unbelonging, nativism, populism, or cultural racism, due to their knowledge of academic literature in the field. This expertise enables them to associate comments with these specific concepts. This observation suggests that when tasking LLMs with analysis through a specific conceptual framework, it is essential to train the model through in-context learning[5] and provide specific examples of concepts we want to be identified. In the evaluation of the classification results, it was particularly noteworthy that the categories identified by GPT-4 remained neutral and did not introduce or enhance any stereotypes present in the comments. While GPT-4 identified broad themes such as stereotypes or prejudice, it refrained from labeling any content as racist. For example, it would point out references to cultural differences between Roma and non-Roma but maintained a very neutral description and avoided labeling the discourse as (cultural) racism (Table 2). It is noteworthy to mention that even though the model repeatedly reminded researchers that the comments contained inflammatory or discriminatory content, it preserved a high level of neutrality when labeling the content.

---

[5] In-context learning is a computational technique used with LLMs where the model learns to perform specific tasks by analyzing examples presented in its immediate input through prompt engineering, rather than through explicit prior fine-tuning on a similar task.

**Table 2: Human categories to describe the dataset.**

| Category | Description |
|---|---|
| Populism: | Populism is a discourse that is used by elites including mainstream media, politicians, and academics to advance political interests in a manner that reflects the alleged will of the "ordinary people", for example, "fears" of uncontrollable immigration, declining economic prosperity, a decline in moral, cultural and religious values, and a loss of national identity and autonomy. |
| Nativism: | Nativism is an exclusionary citizenship discourse constructed around adverse narratives about "'us' (the natives) versus 'them' (the non-natives)," with the latter being perceived as dangerous, as social, economic, or cultural threats to the people of the land. It is a mythicized idea about a disenfranchised group of people, the natives of the land, who themselves appear to be forgotten and suffer the consequences wrought by immigration and mainstreaming of multiculturalism such as higher demographics, lower wages, unemployment, increase in crime, decline in safety, cultural changes and altering of immediate surroundings. |
| Extreme hate speech: | Extreme hate speech encompasses abusive or dehumanizing language invoking well-trodden stereotypes about groups of people, at times endorsing violence, even in playful, humorous ways. |
| Cultural Racism: | Racism is not expressed about biological categories but alludes to culture as a marker of otherness. |
| (Non)Belonging | Roma as the 'other' European who should not be confused with Romanians or Bulgarians. References to their Indian origin and physical resemblance. |
| Perceptions and Stereotypes | Roma as beggars, thieves, scammers and unproductive citizens |

As with the previous categories, we instructed GPT-4 to assign one of the human-defined categories to each comment as a first step and only added the 'None' option in the second step. Surprisingly, with this approach, GPT-4 assigned specific labels to less than half of the posts, with 51.48% (244 samples) being classified as not belonging to any of the assigned categories. This highlights the importance of careful selection of categories and prompt design to ensure accurate and meaningful classification.

**Figure 2: GPT-4 Theme Distribution (Human Categories)**

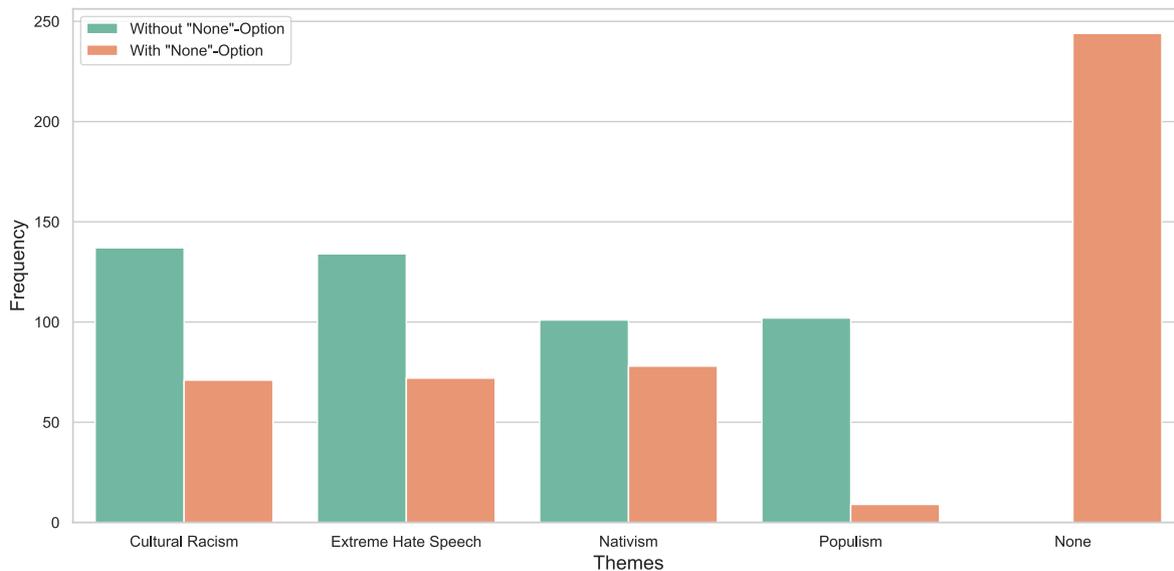

When evaluating the classification, experts agreed that some of the comments did not fit neatly into the categories identified by GPT-4. For example, comments specifically targeting Roma in a derogatory manner were usually labeled as 'Ethnic Misunderstandings and Identity Confusion' or 'Stereotyping and Social Prejudice.' This is one example:

*with the lack of Gypsies in Bulgaria, property prices have been on the rise lately. Keep the Gypsies ... send down the sexy Swedish bikini team to Sunny Beach Bulgaria, where blonds are welcome and Gypsies are not.*

While the experts agreed with most of GPT-4's classifications, some discrepancies were noted. For instance, the comment: *They need to be called GYPSIES...any of them is not Romanian or Bulgarian. The largest community of gypsies from the EU is in Hungary*, which was categorized under 'Cultural Racism' by GPT-4. Experts, however, identified this as hate speech due to the use of a derogatory term that lacks semantic value to denote an ethnic group. The term 'GYPSIES,' particularly in the Eastern European context, is laden with negative connotations such as being unreliable, lazy, or dirty (Breazu, 2020). This language is clearly intended to demean and dehumanize the Roma community and can be labeled as hate speech. Additionally, GPT-4's failure to address the capitalization issue further emphasizes the need for context sensitivity, as the capitalized term 'GYPSIES' indicates an aggressive and derogatory emphasis typical of hate speech. The comment also reflects a discourse of unbelonging, suggesting that Roma should not be associated with or referred to as Romanian or Bulgarian.

Another example, *We will holiday in Sweden no longer. It is not Sweden anymore............*, was classified as 'Nativism' by GPT-4. While the borderline between nativist and populist discourses is very thin (Riedel, 2018) and sometimes hard to distinguish, experts contended that this comment is more indicative of populism rather than nativism. Populism often involves rhetoric that appeals to the 'common people' against a perceived elite or cultural threat (Stavrakakis, 2017) which fits the sentiment expressed in the comment. The statement reflects a broader sense of discontent and nostalgia for an imagined past (in this case, Sweden as a great place for holidays which has lost its appeal because of the presence of Roma beggars), characteristic of populist discourse. This misalignment in thematic understanding highlights a critical gap; GPT-4's classification missed the broader political and cultural context implied by the comment, focusing instead on a narrower interpretation related to native identity versus foreign influence.

These examples illustrate why the human researchers' expertise provided a more detailed and contextualized classification. Their deeper understanding of the socio-political and cultural background allowed them to correctly identify hate speech and the broader populist sentiment, which GPT-4's broader and more generalized categories failed to capture. This shows the importance of incorporating specific examples and context-driven learning when training AI models to analyze complex and sensitive issues like immigration discourse.

**Summary and Conclusion**

This study has explored the potential and limitations of using GPT-4, a state-of-the-art Large Language Model (LLM), to perform thematic analysis (TA) on YouTube comments related to Roma beggars in Sweden. Our experimental study highlights the efficiency and scalability of LLMs in processing large datasets and identifying broad themes within qualitative data. However, it also shows the necessity for human oversight to ensure depth, context, and accuracy in qualitative research.

*Human-AI Synergy in Qualitative Analysis*

The integration of LLMs in qualitative research presents promising opportunities for enhancing research methodologies, but it also introduces significant challenges. As Byun et al. (2023) suggest, AI can match human capabilities in processing qualitative data, but it requires careful guidance to avoid discrepancies between AI and human reasoning. Our study demonstrates that while GPT-4 can generate useful initial categorizations, the depth and specificity provided by human researchers are crucial for accurate and meaningful analysis. The broader, neutral approach of GPT-4 often fails to capture the thorough understanding that human expertise brings to thematic analysis (Bano et al., 2024).

*Context Learning*

One of the critical insights from this study is the importance of context learning in deploying LLMs for qualitative research. GPT-4's ability to process and categorize data can be significantly improved by training the model with specific contextual information. Providing the model with a detailed background and socio-political context allows it to produce more accurate and relevant classifications. This step is essential for refining the model's capabilities to ensure it can effectively interpret and analyze complex qualitative data (Gao et al., 2023; De Paoli, 2023).

*Theory-Driven Prompts*

Another key finding is the effectiveness of theory-driven prompts in guiding LLMs. Using predefined theoretical frameworks and detailed task descriptions helps LLMs like GPT-4 align more closely with human categorizations (Mu et al., 2023). This approach leverages the strengths of LLMs in processing large volumes of data while ensuring that the analysis adheres to established academic theories and frameworks (Kennedy & Thornberg, 2018). By incorporating specific examples and theoretical concepts into the prompts, researchers can improve the model's performance and reduce the risk of misclassification (Chen et al., 2023).

*Discrepancies and Refinement*

The discrepancies observed in the thematic analysis indicate a need for further refinement in GPT-4's understanding of context and subtle language. Teaching the model context and providing it with sufficient background information are crucial steps to improve its accuracy. Currently, we are employing a top-down (deductive) approach, where the model operates with predefined beliefs and normative values, which introduces its own subjectivity. To reduce this subjectivity, it is essential to refine the model's analysis through context-specific training and clearer methodological steps. By instructing the model about the context of the comments and feeding it relevant theoretical frameworks, we can enhance its ability to provide more accurate and meaningful classifications.

The two category schemes reveal distinct approaches to categorizing immigration discussion themes. The first uses broader, neutral terms like 'Cultural Clash and Integration Challenges,' while the second employs more specific, potentially controversial labels such as 'Cultural

Racism' and 'Extreme Hate Speech.' Despite these differences, both graphs show a similar pattern: the introduction of a 'None' option significantly alters response distribution. Without the 'None' option, responses spread across available categories. However, when 'None' is introduced, it becomes the dominant choice, suggesting many responses don't fit neatly into predefined themes. This shift is more pronounced in the second graph, where provocative labels may have pushed more responses toward 'None.' The comparison highlights how category selection and the inclusion of a 'None' option impact data interpretation. The first graph's neutral categories might lead to less charged discussions, while the second's terminology could prompt more contentious debates. In both cases, the high frequency of 'None' responses underscores the complexity of immigration discourse and the limitations of rigid categorization.

*Future Directions*

Looking ahead, future research should focus on further refining the synergy between human intelligence and LLM capabilities. This involves developing more sophisticated methods for integrating in-context learning and theory-driven prompts into the training and deployment of LLMs. Additionally, it is crucial to address the ethical considerations and limitations associated with using LLMs, particularly in handling sensitive data and ensuring the accuracy and reliability of the analysis (Polonsky & Rotman, 2023; Bano et al., 2024).

Moreover, exploring ways to improve the interpretative abilities of LLMs and incorporating feedback mechanisms where human researchers can interactively refine and guide the model's analysis will be vital. This collaborative approach can use the strengths of both AI and human expertise, which can lead to more thorough and comprehensive qualitative research outcomes (De Paoli, 2023; Rudolph et al., 2023).

While LLMs like GPT-4 hold significant potential for transforming qualitative analysis, their deployment must be carefully managed to ensure they complement rather than replace human expertise. Our findings emphasize this need: although GPT-4 can quickly process large datasets and identify broad themes, it often misses the contextual insights that human expertise provides. For example, GPT-4's broader, neutral approach sometimes led to the misclassification of comments, while human researchers, with their deep understanding of socio-political contexts and academic literature, could identify specific themes more accurately. Furthermore, the AI's tendency to avoid explicitly labeling hate speech highlighted the need of human oversight to interpret and categorize sensitive data correctly. The future of qualitative research could definitively benefit from a synergistic approach that combines the scalability and efficiency of AI with the critical thinking and contextual understanding of human researchers.